\documentclass{ws-p8-50x6-00}

\usepackage{latexsym}
\input{psfig}
\def\Ga    {g_{_{A}}}
\def\Gv    {g_{_{V}}}

\def\Za    {Z_{_A}}
\def\Zv    {Z_{_V}}
\begin{document}

\title{Lattice study of nucleon properties with domain wall fermions
}

\author{Shoichi Sasaki}

\address{Department of Physics, University of Tokyo,\\
7-3-1 Hongo, Tokyo 113-0033, Japan\\
E-mail: ssasaki@phys.s.u-tokyo.ac.jp}

\maketitle

\abstracts{
Domain wall fermions (DWF) are a
new fermion discretization scheme with greatly improved 
chiral symmetry. Our final goal is 
to study the nucleon spin structure through lattice simulation 
using DWF. In this paper, we present our current progress 
on two topics toward this goal: 1) the mass spectrum of the nucleon excited states
and 2) the iso-vector vector and axial charges, $g_{V}$ and 
$g_{A}$, of the nucleon.}

\noindent
{\bf 1. Introduction}
\vspace{0.2cm}

The RIKEN-BNL-Columbia-KEK QCD Collaboration has been 
pursuing the domain-wall fermion (DWF) method\cite{DWF} in 
lattice quantum chromodynamics (QCD).  
In DWF an extra fifth dimension is added to the lattice.  
By manipulating the {\it domain-wall} structure of the fermion mass
in this fifth dimension, we control the number of light fermion species
in the other four space-time dimensions. These light fermions 
possess exact chiral symmetry in the limit of an infinite fifth dimension. 
In particular: 1) fermion near-zero mode effects are well 
understood\cite{rbc00},
2) explicit chiral symmetry breaking induced by a finite extra dimension 
is described by a single residual mass parameter, which is very small in the
present calculation, in the low-energy effective
lagrangian\cite{rbc00}, and  3)
non-perturbative renormalization works well\cite{rbc01}.

DWF is a promising new approach for treating fermions on the
lattice. However, we need several tests of DWF in the baryon sector 
to reach our final goal of establishing the spin structure of the nucleon from
first principles. 
Here we report our recent studies 
of the mass spectrum of the nucleon and its excited states and 
the nucleon matrix elements of the iso-vector vector and axial charges
in quenched lattice QCD with DWF. Although most of the latter 
results are preliminary, the conclusive results in 
the former subject have been reported in
Ref.[4]. 

\vspace{0.2cm}
\noindent
{\bf 2. Nucleon excited states}
\vspace{0.2cm}

First, we discuss the mass
spectrum of the nucleon $N$ and its excited states
(the negative-parity nucleon $N^*$ and the positive-parity
first excited nucleon $N'$) by means of a systematic investigation 
utilizing two distinct interpolating operators 
$B_{1}^{\pm}$ and $B_{2}^{\pm}$. For an explanation of those operators,
see Ref.[4].
Our quenched DWF calculation was employed on lattice with size 
$16^3 \times 32 \times 16$, gauge coupling $\beta = 6/g^2 = 6.0$, and
domain wall height $M_{5}=1.8$. Additional details of our 
simulation can be found in Ref.[4].

In Fig.1 we show the low-lying nucleon spectrum 
as a function of the quark mass, $m_{f}$
in lattice units ($a^{-1}\approx 1.9$ GeV set from 
$aM_{\rho}$=0.400(8) in the chiral limit).
$B_{1}^{+}$ gives the ground-state nucleon mass $N$ (cross). 
The $N^*$ mass estimates (square and diamond) are extracted 
from both $B_{1}^{-}$ and $B_{2}^{-}$. The corresponding 
experimental values for $N$ and $N^*$ are marked with 
lower and upper stars. Both $N^*$ mass estimates extracted from 
two distinct operators agree with each other.
The large $N$-$N^*$ mass splitting is clearly evident. 

In contrast to the negative parity operators,
we find that the mass estimates from a 
second operator, $B_{2}^{+}$, are considerably larger than 
the ground state obtained from $B_{1}^{+}$. This suggests that $B_{2}^{+}$ has
negligible overlap with the nucleon ground state and
provides a signal for the positive-parity {\it excited}
nucleon $N'$. To justify this possibility, we employ a sophisticated 
approach which utilizes the transfer matrix of a $2\times 2$ 
correlation function\cite{diag} constructed from both $B_{1}^{+}$ and $B_{2}^{+}$. 
The diagonalization of the transfer matrix yields the excited state\cite{diag}.
Fig.2 shows a comparison of the fitted mass from 
$\langle\langle B_{2}^{+}(t){\bar B}_{2}^{+}(0)\rangle\rangle$ 
(circle) and the estimated mass from the average effective mass given by
the smaller eigenvalue of the transfer matrix (bullet).
The cross symbol corresponds to the nucleon ground state
mass evaluated from the larger eigenvalue of the transfer matrix,
which is quite consistent with the fitted mass from 
$\langle\langle B_{1}^{+}(t){\bar B}_{1}^{+}(0)\rangle\rangle$ 

Another important conclusion can be drawn from Fig.1 and Fig.2.
In the heavy quark mass region,
the ordering of the negative-parity nucleon
($N^*$) and the positive-parity excited nucleon ($N'$) 
is inverted relative to experiment. This remarkable
result was originally reported in our early paper\cite{sbo01} and 
subsequently confirmed in Ref.[6]. Further
systematic calculation is required to determine whether this ordering
switches to the observed ordering as one approaches the chiral limit.

%
%
\begin{figure}[t]
\begin{center}
\epsfxsize=7.5cm 
\epsfbox{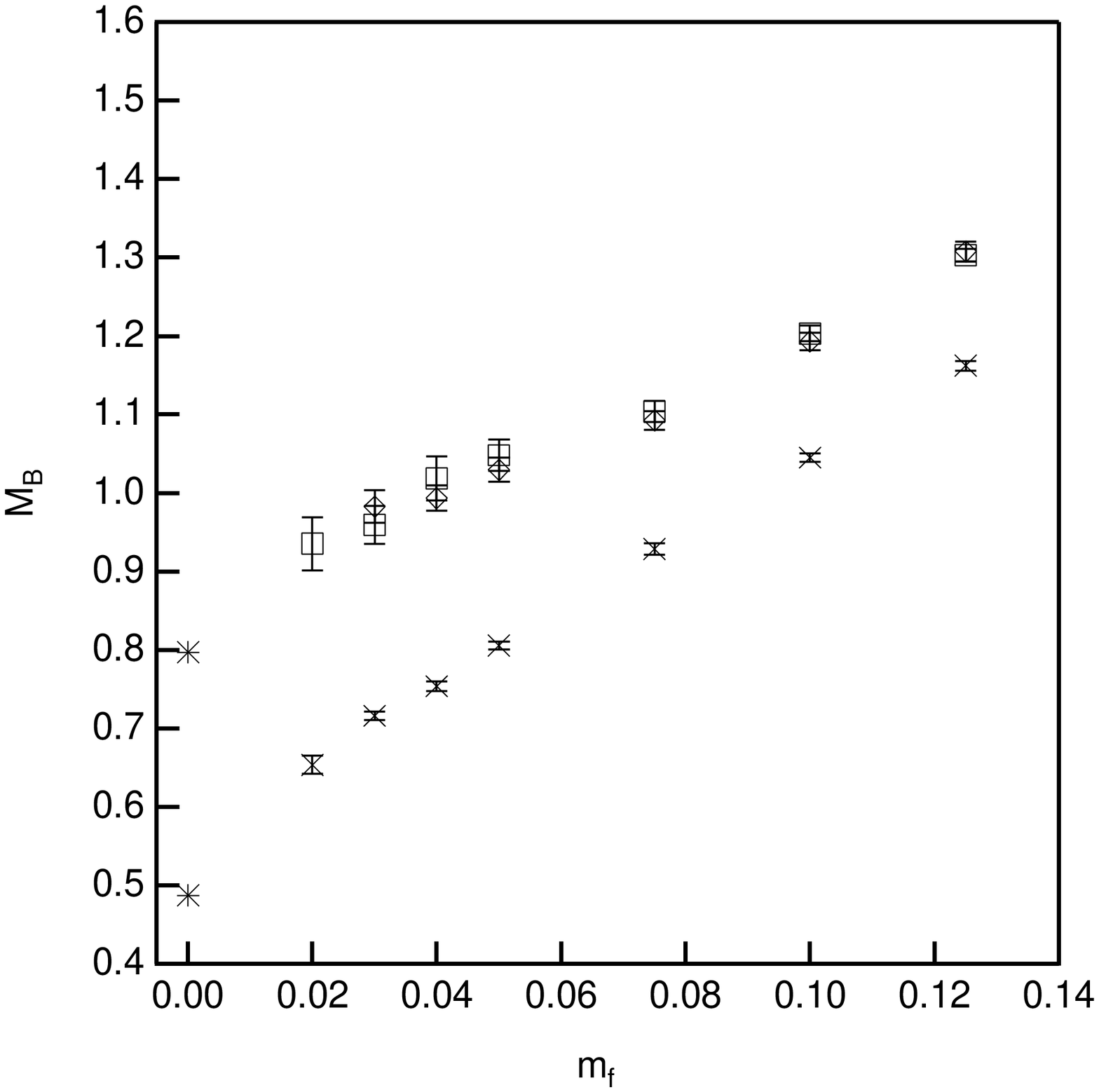} 
\caption{\label{fig:neg-spect}
$N$ and $N^*$ (square and diamond) masses versus the quark mass
$m_{f}$ in lattice units. Note 
the large $N$-$N^*$ mass splitting which is within 10\% (in the chiral
limit) of the experimental value (bursts).}
\epsfxsize=7.5cm 
\epsfbox{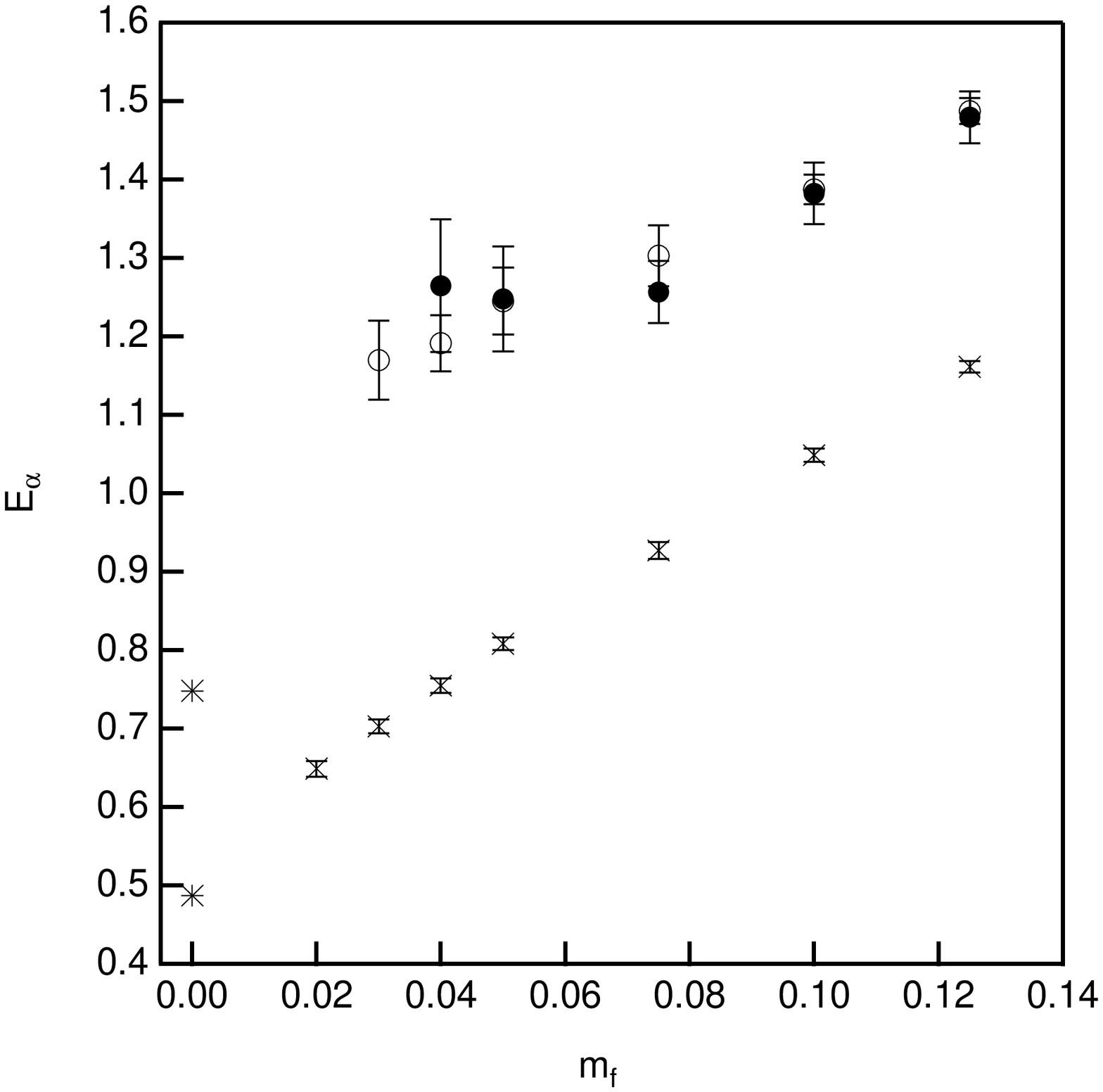} 
\caption{\label{fig:pos-spect}
The mass of the positive-parity excited state (circles) is too high compared
to the nucleon ground state (cross) to account for the observed splitting.
}
\end{center}
\end{figure}

\vspace{0.2cm}
\noindent
{\bf 3. Nucleon matrix elements}
\vspace{0.2cm}

The nucleon (iso-vector) axial charge $\Ga$ 
is a particularly interesting quantity. We know precisely the experimental 
value $\Ga=1.2670(35)$ from neutron beta decay. Why does
$\Ga$ deviate from unity in contrast to vector charge, $\Gv=1$?
The simple explanation is given by 
the fact that the axial current is only partially conserved in 
the strong interaction while the vector current is exactly conserved. 
Thus, the calculation of $\Ga$ is an especially relevant test of
the chiral properties of DWF in the baryon sector.
In addition, calculation of $\Ga$ is an important first step 
in studying polarized nucleon structure functions since
$\Ga=\Delta u - \Delta d$ where 
$\langle p,s|{\bar q_{f}}\gamma_{5}\gamma_{\mu}q_{f}|p,s\rangle
=2s_{\mu}\Delta q_{f}$ with $s^2=-1$ and $s\cdot p=0$.

We follow the standard practice\cite{kek,desy} for the
calculation of $\Gv$ and $\Ga$. 
The three-point functions for the 
local vector current  $J_{_{V}}^{f}={\bar q}_{f}\gamma_{4}q_{f}$ 
and axial current
$J_{_{A}}^{f}={\bar q}_{f}\gamma_{5}\gamma_{i}q_{f}$ ($i=1,2,3$) are 
defined by
%
%
\begin{equation}
G_{\Gamma}^{u, d}(t,t')={\rm Tr}[P_{_\Gamma}\sum_{{\vec x},{\vec x}'}
\langle TB_{1}(\vec x,t)J_{_{\Gamma}}^{u,d}({\vec x}',t'){\bar B}_{1}(0,0)\rangle]
\end{equation}
with $\Gamma$ = $V$ (vector) or $A$ (axial) where 
$P_{_{V}}=P_{+}=(1+\gamma_{4})/2$ and 
$P_{_{A}}=P_{+}\gamma_{i}\gamma_{5}$.
For the axial current, the three-point function is averaged over 
$i=1,2,3$. The lattice estimates of vector and axial charges can
be derived from the ratio between two- and three-point functions
%
%
\begin{equation}
g_{_{\Gamma}}^{\rm 
lattice}={{G_{\Gamma}^{u}(t,t')-G_{\Gamma}^{d}(t,t')}
\over G_{_{N}}(t)}\;,
\end{equation}
where $G_{_{N}}(t)={\rm Tr}[P_{+}\sum_{\vec x}\langle T B_{1}({\vec 
x},t){\bar B}_{1}(0,0)\rangle]$.

Recall that in general
lattice operators ${\cal O}_{\rm lat}$ and continuum operator
${\cal O}_{\rm con}$ are regularized in different schemes. The 
operators are related by a renormalization factor $Z_{\cal O}$:
${\cal O}_{\rm con}(\mu)=Z_{\cal O}(a\mu){\cal O}_{\rm lat}(a)$. 
This implies that the continuum value of vector and axial charges
are given by $g_{_{\Gamma}}=Z_{_{\Gamma}}g_{_{\Gamma}}^{\rm 
lattice}$. In the case of conventional Wilson fermions, the 
renormalization factor $\Za$ is usually estimated in  
perturbation theory ($\Za$ differs from unity because 
of explicit symmetry breaking). For DWF, the conserved axial current 
receives no renormalization. This is not true for the lattice local 
current. An important advantage with DWF, however, is that the 
lattice renormalizations, $\Zv$ and $\Za$, 
of the local currents are the same\cite{rbc01} 
so that the ratio 
$(\Ga/\Gv)^{\rm lattice}$ directly yields the continuum value
$\Ga$.

Our preliminary results are analyzed on 200 quenched gauge 
configurations at $\beta=6.0$ on a $16^3 \times 32 \times 16$ 
lattice with $M_{5}=1.8$\cite{sbo00}. 
We choose a fixed separation in time of the
nucleon interpolating operators,
$t=t_{\rm source}-t_{\rm sink}$ and $t' < t$,
with currents inserted in between. We take $t_{\rm source}=5$ and
$t_{\rm sink}=21$\cite{sbo00}.

In Fig.3 we show the dependence of the vector renormalization, 
$\Zv=1/\Gv^{\rm lattice}$ on the location of current insertions.
A good plateau is observed so that $\Zv$ is certainly well behaved.
The value 0.763(5) at $m_{f}=0.02$ (obtained by averaging 
over time slices denoted by the solid line
in Fig.3) agrees well with 
$\Za=0.7555(3)$,
which was obtained from a completely different calculation of meson two-point
correlation functions based on
the relation
$\langle A_{\mu}^{\rm conserved}(t){\bar q}\gamma_{5}q(0)\rangle=
\Za\langle A_{\mu}^{\rm local}(t){\bar q}\gamma_{5}q(0)\rangle$\cite{rbc00}. 

For the axial charge, $\Ga^{\rm lattice}$, 
plateaus are seen for $10 \leq t \leq 16$ in Fig.4, so
the charge ratios $(\Ga/\Gv)^{\rm lattice}$ at each $m_{f}$ are 
averaged in this time slice range. We find that there is a fairly strong 
dependence of $(\Ga/\Gv)^{\rm lattice}$ on $m_{f}$ as be shown in Fig.5.
A simple linear extrapolation to $m_{f}=0$ 
yields $\Ga=0.62(13)$, which is roughly 1/2 the 
experimental value\cite{sbo00}. However, a simple linear ansatz
may not describe the data in which case the result in the
chiral limit may be even smaller.

To compare to results using Wilson fermions,
we plot $\Ga$ versus the square of the $\pi$-$\rho$ mass
ratio in Fig.6. Our result is given by the bullets.
We also include two heavier mass points (with large errors)
from an earlier simulation 
using a larger separation between the source and the sink.
The triangles and diamonds are from Wilson simulations at relatively 
strong\cite{kek} and weak coupling\cite{desy}. 
At first glance, the DWF and Wilson fermions seem to be 
in rough agreement except for the 
lightest point. However, our DWF results 
have a strong mass dependence.
This may be a 
finite volume effect; the  
Wilson results at strong coupling were simulated on
a lattice with a volume which is twice ours.

A couple of comments on the mass dependence of our data are in order. 
First, it is interesting to note that our results appear
consistent with
the value 5/3 (marked as star) in the heavy quark limit, 
while the others seem inconsistent with this limit. 
Second, our results may also be consistent with vanishing
in the chiral limit. This can be 
explained through the axial Ward-Takahashi identity which governs $\Ga$.
If the PCAC relation $m_{\pi}^2 \propto m_{f}$ is
modified, for example by chiral logarithms, 
the nucleon matrix element of the pseudoscalar 
density does not develop a pole as $m_{f}\rightarrow 0$,
and the r.h.s of the identity vanishes in the chiral limit. 
Thus, $\Ga$ must also vanish as $m_{f}\rightarrow 0$.
Indeed, we already know that
the PCAC relation for the pion mass is modified in the quenched 
approximation by two effects: zero-modes of the Dirac operator
and the quenched chiral logarithm\cite{rbc00}. Further investigation of
the Ward-Takahashi identity is under way. 

%
%
\begin{figure}[t]
\begin{center}
\epsfxsize=7.5cm 
\epsfbox{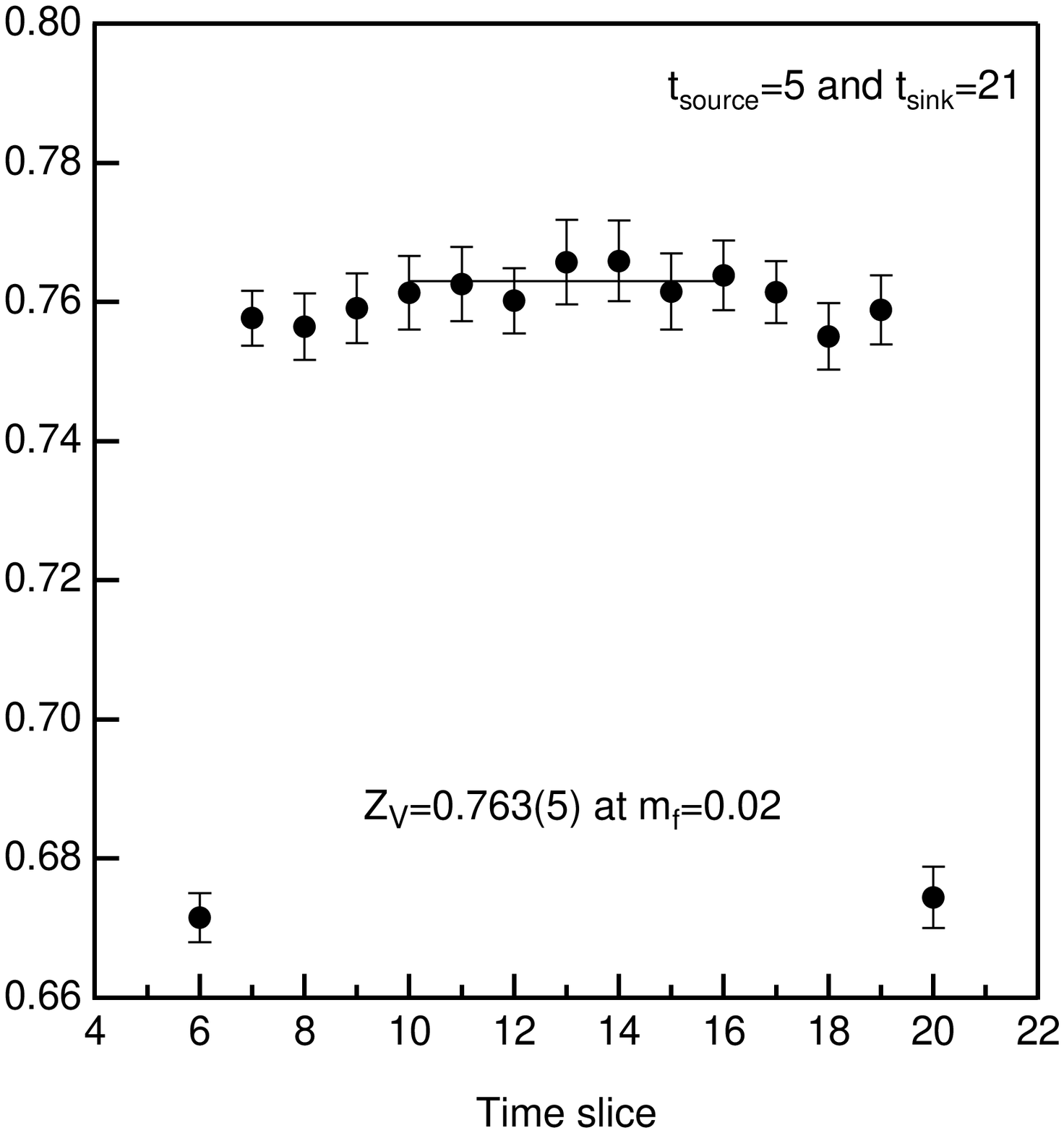} 
\caption{\label{fig:Zv}
Dependence of vector renormalization, $Z_{V}=1/g_{V}^{\rm lattice}$,
on $t'$, at $m_{f}=0.02$. A good plateau is observed.}
%
%
\epsfxsize=7.5cm 
\epsfbox{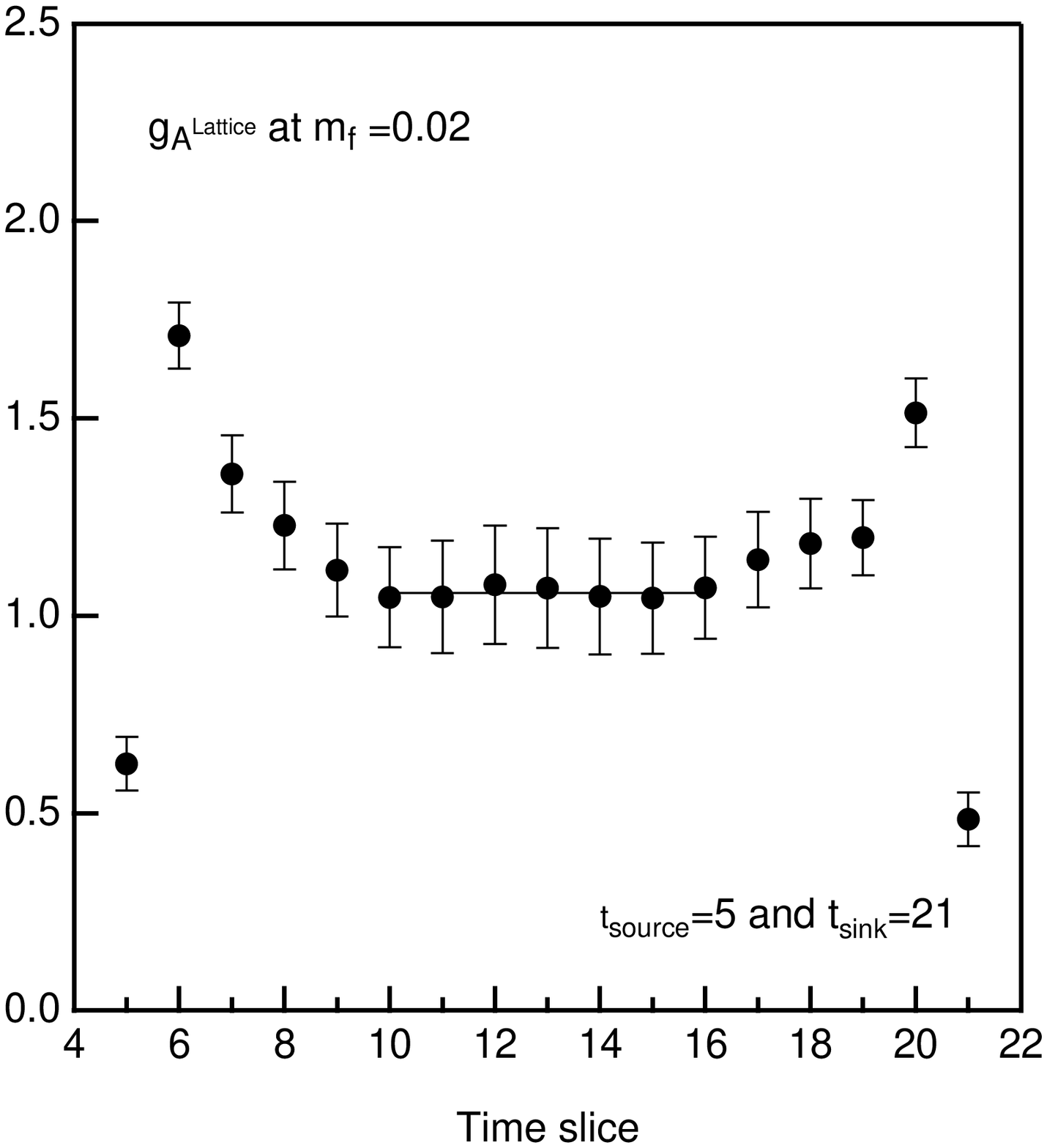} 
\caption{\label{fig:LGa}
The lattice axial charge, $g_{A}^{\rm lattice}$, at $m_{f}=0.02$.
A good plateau is seen in the range $10 \leq t \leq 16$.
}
\end{center}
\end{figure}
%

%
%
\begin{figure}[t]
\begin{center}
\epsfxsize=7.5cm 
\epsfbox{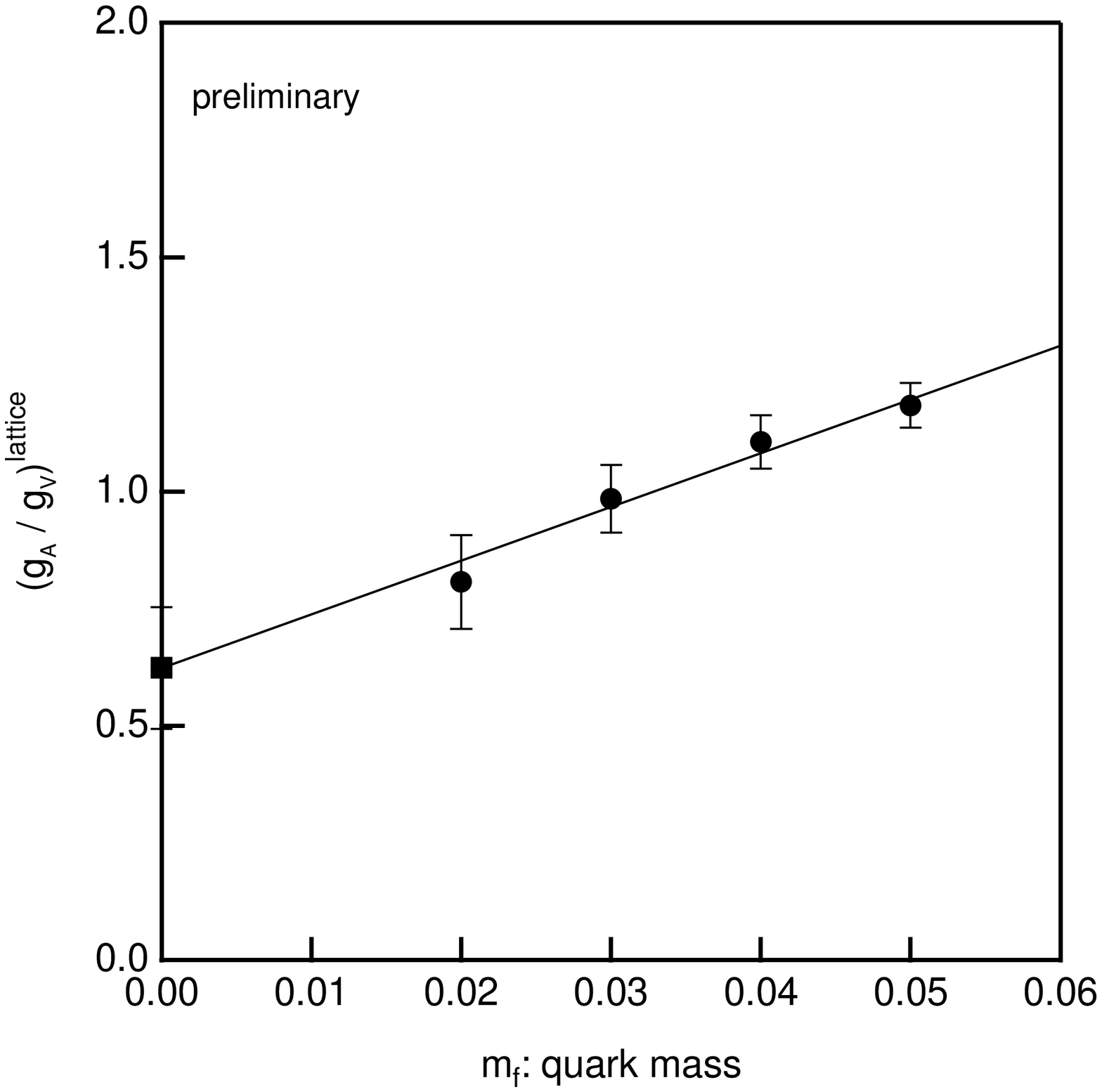} 
\caption{\label{fig:Ga-ave}
Dependence of $(g_{A}/g_{V})^{\rm lattice}$ on $m_{f}$.}
%
%
\epsfxsize=7.5cm 
\epsfbox{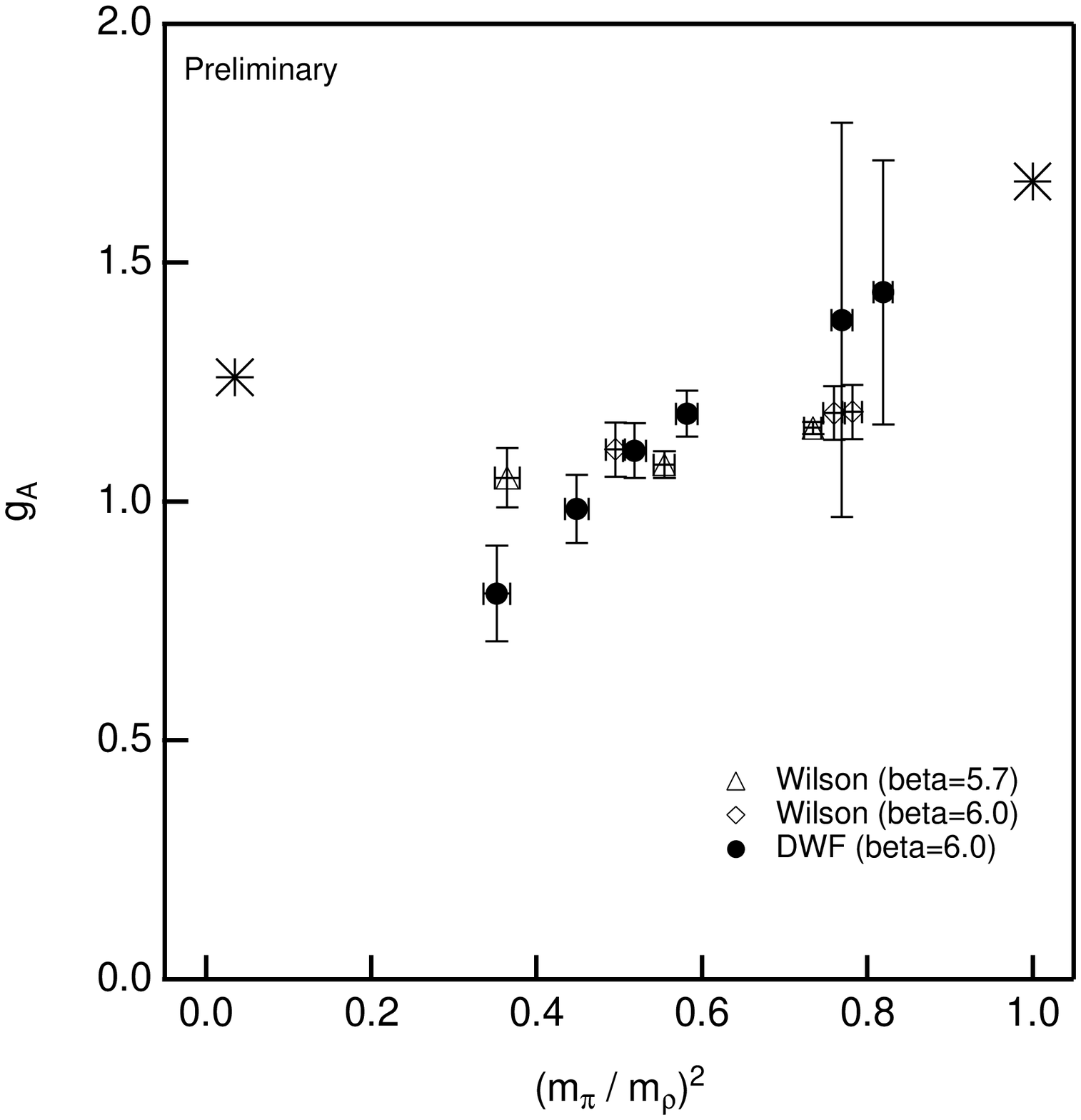} 
\caption{\label{fig:ratio-ga}
$g_{A}$ versus $(m_{\pi}/m_{\rho})^2$.}
\end{center}
\end{figure}

\vspace{0.2cm}
\noindent
{\bf 4. Conclusions}
\vspace{0.2cm}

We have explored several nucleon properties in quenched lattice QCD
using domain wall fermions toward our final goal of studying the
spin structure of the nucleon from first principles.

Our quenched DWF calculation reproduces 
very well the large mass splitting between the nucleon
$N(939)$ and its parity partner $N^*(1535)$\cite{sbo01}. 
We have also calculated the mass of the first positive-parity
excited state $N'(1440)$ by the diagonalization
of a $2\times 2$ matrix correlator and confirmed that it is heavier than 
the negative-parity excited state $N^*(1535)$\cite{sbo01}.
A remaining puzzle is whether or not a switching of 
$N^*$ and $N'$ occurs close to the chiral limit.

A preliminary calculation of iso-vector vector and axial charges shows
that all the relevant three-point functions are well behaved. 
$\Zv$ determined from the nucleon matrix element of the vector current 
agrees closely with that from an NPR study of quark 
bilinears\cite{rbc01} and
a direct calculation using meson correlation functions\cite{rbc00}. 
This indicates that $\Gv=1$ and $\Zv=\Za$ are mutually satisfied in
our quenched DWF calculation. However, a linear extrapolation
of $\Ga$ to the chiral limit yields a value which is a factor of
two smaller than the experimental value. 
We are currently investigating the Ward-Takahashi identity that
governs $\Ga$ to shed light on this behavior.
We also plan to check related 
systematic effects arising from finite lattice volume and quenching
(for example quenched chiral logarithms, zero modes, and the absence of
the full pion cloud), especially
in the lighter quark mass region.

\vspace{0.2cm}
\noindent
{\bf Acknowledgments}
\vspace{0.2cm}

The author would like to thank the organizers of NSTAR2001 for an
invitation. This work was done in collaboration with 
Tom Blum and Shigemi Ohta as a part of the RIKEN-BNL-Columbia-KEK QCD 
Collaboration. We thank RIKEN, Brookhaven National Laboratory, and 
the U.S. Department of Energy for providing the facilities essential 
for the completion of this work.

\end{document}